\documentclass[aps,prl,twocolumn,eqsecnum,superscriptaddress,floatfix,10pt]{revtex4-2}
\usepackage{amsmath,amssymb,amsfonts,stmaryrd,wasysym,graphicx,multirow,color,textcomp,subfigure,nicefrac,float,enumitem}
\usepackage{url}
\usepackage[colorlinks=true,citecolor=blue,urlcolor=blue,linkcolor=blue]{hyperref}

\newcommand{\cyan}[1]{{\color{cyan}{\it {#1}}}}

\def\be{\begin{equation}}
\def\ee{\end{equation}}
\def\bea{\begin{eqnarray}}
\def\eea{\end{eqnarray}}

\counterwithout{equation}{section}

\begin{document}

\title{$\mathcal{PT}$-symmetric Non-Hermitian Hopf Metal}
\author{Seik Pak}
\address{Department of Physics, Hanyang University, Seoul 04763, Republic of Korea}
\author{Cheol Hun Yeom}
\address{Department of Physics, Hanyang University, Seoul 04763, Republic of Korea}
\address{Department of Physics, Konkuk University, Seoul 05029, Republic of Korea}
\author{Sonu Verma}
\email{sonu.vermaiitk@gmail.com}
\affiliation{Center for Theoretical Physics of Complex Systems, Institute for Basic Science, Daejeon 34126, Korea}
\author{Moon Jip Park}
\email{moonjippark@hanyang.ac.kr}
\address{Department of Physics, Hanyang University, Seoul 04763, Republic of Korea}
\date{\today}

\begin{abstract}
Hopf insulator is a representative class of three-dimensional topological insulators beyond the standard topological classification methods based on $K$-theory. In this letter, we discover the metallic counterpart of the Hopf insulator in the non-Hermitian systems. While the Hopf invariant is not a stable topological index due to the additional non-Hermitian degree of freedom, we show that the $\mathcal{PT}$-symmetry stabilizes the Hopf invariant even in the presence of the non-Hermiticity. In sharp contrast to the Hopf insulator phase in the Hermitian counterpart, we discover an interesting result that the non-Hermitian Hopf bundle exhibits the topologically protected non-Hermitian degeneracy, characterized by the two-dimensional surface of exceptional points. Despite the non-Hermiticity, the Hopf metal has the quantized Zak phase, which results in bulk-boundary correspondence by showing drumhead-like surface states at the boundary. Finally, we show that, by breaking $\mathcal{PT}$-symmetry, the nodal surface deforms into the knotted exceptional lines. Our discovery of the Hopf metal phase firstly confirms the existence of the non-Hermitian topological phase outside the framework of the standard topological classifications.
\end{abstract}

\maketitle


\cyan{Introduction--} In three-dimensions, the insulating phases with two-band systems are topologically classified by Hopf invariant, where the non-trivial topology is characterized by the eigenspinors with the Hopf bundle structure \cite{HopfGangWen,HopfCenke,HopfDuan,Hopf1,Hopf2}. The Hopf insulator has gathered great interest due to its unusual phenomena. That is, the topological stability only persists when the number of the band is equal to two. The addition of auxiliary trivial bands trivializes the topological index. This instability features incompatibility of  \textit{K}-theory-based classification methods for the description of the Hopf insulator \cite{RevModPhys.88.035005,Ryu_2010,PhysRevB.78.195125,10.1063/1.3149495}. In this regard, there has been a considerable effort in studying such delicate topological phases both experimentally \cite{PhysRevLett.130.057201} and theoretically \cite{PhysRevB.106.075124}. As a similar phenomenon but with different physical origins, the fragile topology has also been suggested as an interesting phase that is detectable by the symmetry indicator even with the trivial \textit{K}-theory index \cite{PhysRevLett.121.126402,FT1,FT2,FT3,FT4,FT6,FT7,FT5,FT8,FT9,FT10}.

Non-Hermitian systems have emerged as a transformative area to extend the knowledge of topological phases. In general, the complex energy spectra can enrich or nullify the Hermitian topological classifications \cite{PhysRevX.9.041015,PhysRevB.100.081104,NH1,NH2,NH3,NH4,NH5,NH6,NH7,NH8,NH9,NH10,NH11,NH12}. Meanwhile, in symmetry-protected topological phases, $\mathcal{PT}$ symmetry plays an important role since it ensures either real or complex pairs of eigenenergies. Unlike the Hermitian systems, the
transition from real to complex spectrum occurs by passing through the non-Hermitian degeneracy, known as an exceptional point
(EP) \cite{realHamiltonian,EP1,EP2,EP3,EP4,EP5,EP6,EP7,EP8,EP9,EP10}. While \textit{K}-theory is applicable to the topological classification of EP as well \cite{PhysRevX.9.041015}, the energy spectrum on EP exhibits eigenstates coalescence, which shows completely different physical phenomena from the Hermitian systems.

In this work, we discover the Hopf metal phase, the non-Hermitian counterpart of the Hopf insulator. In general, the introduction of the non-Hermiticity can nullifiy the homotopy classifications of the Hopf bundle. However, we show that the presence of the additional $\mathcal{PT}$ symmetry allows us to define the non-Hermitian Hopf invariant. Unlike the Hopf insulator in the Hermitian phase, the Hopf metal is characterized by the topologically protected two-dimensional surfaces of the EP (exceptional surface), which shows intriguingly different behavior from the Hermitian Hopf insulators. Similar to the other topological systems protected by the finite homotopy group, the addition of the trivial band destabilizes the Hopf metal phase and the exceptional surfaces. In addition, we show that the Hopf metal phase exhibits bulk-boundary correspondence, where the surface state manifests as the drumhead surface state. The addition of $\mathcal{PT}$ symmetry-breaking perturbation results in the deformation of the exceptional surface into the nodal lines of the exceptional points with non-trivial linking structures. Finally, we discuss the experimental realization of the Hopf metal in various experimental platforms. Our work contributes to the understanding of topological phases of matter by firstly discovering the non-Hermitian topological phase protected by Hopf invariant.

\begin{figure*}[t!]
   \centering
    \includegraphics[width=1\linewidth]{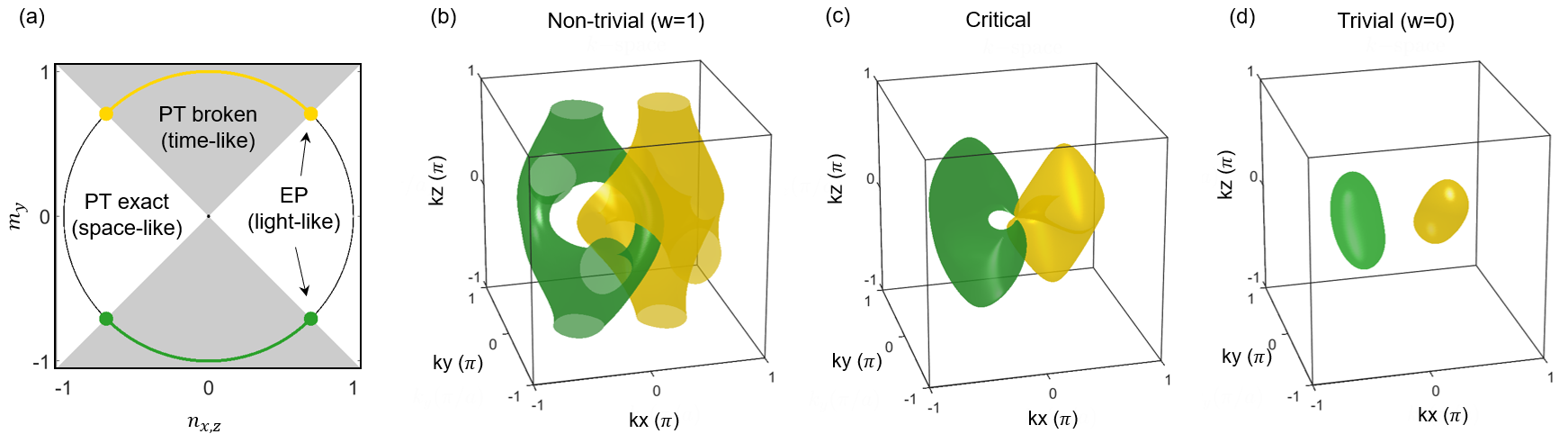}
    \caption{(a) Illustration of $\mathcal{PT}$ symmetric energy spectra. Grey and white region indicates time-like and space-like eigenspinors, where the eigenenergies are purely imaginary and purely real respectively. The boundary between the two regions corresponds to the light cone where the light-like spinors give rise to the exceptional point (green and yellow dots). (b)-(d) the preimage of the light-like spinors (EP) in the Brillouin zone for the case with the  (b) non-trivial ($w=1, m=2$) (c) critical ($m=3$) (d) trivial ($w=0, m=4$) Hopf invariant. The interior of the exceptional surface corresponds to the light-like region, while the exterior is the space-like region. In the non-trivial case, the two exceptional surfaces are linked together. At the critical point, the two exceptional two surfaces intersect. For the trivial case, two nodal surfaces are completely separated. As increasing the value of $m$ further, the surfaces become flatter until they completely disappear.}
       \label{fig1}
\end{figure*}

\cyan{PT-symmetric Non-Hermitian system--} In Hermitian systems, the Bloch Hamiltonian of two-band insulator can be represented by the spinor $\vec{n}(\mathbf{k})$ on the Bloch sphere ($S^2$) as, $h_0(\mathbf{k})=\vec{n}(\mathbf{k})\cdot \vec{\sigma}$, where $\vec{\sigma}$ is the Pauli matrices representing the sublattice (internal) degree of freedom. When the Chern number vanishes in the two-dimensional subspace of the Brillouin zone ($T^3$), the Hopf invariant  ($w (\vec{n})\in \mathbb{Z}$) topologically classifies the two-band insulators \cite{HopfGangWen}. Geometrically, for a given spin  $\vec{n}_0$ in the Bloch sphere, a preimage $\mathbf{k}$ in the BZ such that $\vec{n}(\mathbf{k})=\vec{n}_0$ forms a closed contour. The Hopf invariant counts the linking numbers between the preimages of two different spins. To be explicit, for the spinor configuration $\vec{n}(\mathbf{k})$, the Hopf invariant can be evaluated as\cite{HopfCenke}, 
\begin{align} 
    w (n_x(\mathbf{k}),n_y (\mathbf{k}),n_z (\mathbf{k})) =  \int \frac{d^3k}{(2\pi)^3} \ N^a \partial_{k_x} N^b
\partial_{k_y} N^c
\partial_{k_z} N^d,
\end{align}
where the normalized eigenspinor of $h_0(\mathbf{k})$ is given as, $z(\textbf{k})$ is written as, $z(\bold k) = (
    N_1(\bold k) + iN_2(\bold k), N_3(\bold k) + i N_4(\bold k))^T. $ The non-zero Hopf invariant characterizes the Hopf insulator. The change in the Hopf invariant can only occur by undergoing gapless transitions.

However, as we introduce the non-Hermiticity, the Hamiltonian can generally contain additional anti-Hermitian components as follows. 
\bea
    h(\textbf{k}) &=&  (n_x(\textbf{k})+im_x(\textbf{k}))\sigma_x + (n_z(\textbf{k})+im_z(\textbf{k}))\sigma_z
    \nonumber
    \\
    &+& (n_y(\textbf{k})+im_y(\textbf{k}))\sigma_y,
    \label{Eq:ham}
\eea
where $n_i(\mathbf{k})$ ($m_i(\mathbf{k})$) represent the Hermitian (anti-Hermitian) components. Since the Hamiltonian is now represented by six real parameters \footnote{Here, we ignore the component of the identity matrix since it only shifts the overall energy}, the non-Hermiticity can destabilize the Hopf invariant. Nevertheless, the presence of the additional $\mathcal{PT}$-symmetry, $\mathcal{PT}h(\textbf{k})\mathcal{PT}^{-1} = h(\textbf{k})^*$ \cite{realHamiltonian}, allows us to choose a gauge such that the Hamiltonian can be represented as a real-valued matrix. Accordingly, three of the parameters vanish ($n_y(\mathbf{k})=m_x(\mathbf{k})=m_z(\mathbf{k})=0$). The remaining non-zero Hermitian and anti-Hermitian parameters $\big(n_x(\mathbf{k}),m_y(\mathbf{k}),n_z(\mathbf{k}) \big)$ form a three-component vector that can be projected on $S^2$. Therefore,  $\mathcal{PT}$-the symmetric non-Hermitian system can be classified by the modified Hopf invariant, $w (n_x(\textbf{k}),m_y(\textbf{k}),n_z(\textbf{k}))$, which is well-defined except for transition point characterized by null Hamiltonian $h(\mathbf{k})=0$.

In the complex energy plane, the energy dispersion of Eq. \eqref{Eq:ham} is generally complex-valued, which is given as, 
\bea
    E(\textbf{k}) &= \pm\sqrt{|\mathbf{n}(\textbf{k})|^2-|\mathbf{m}(\textbf{k})|^2+2i\mathbf{n}(\textbf{k})\cdot \mathbf{m}(\textbf{k})}.
    \label{eq:PTEnergy}
\eea
The presence of $\mathcal{PT}$-symmetry ensures that the Hermitian and anti-Hermitian vectors are perpendicular to each other, $\vec{n}(\mathbf{k})\perp \vec{m}(\mathbf{k})$. The corresponding energy eigenvalues are confined to either real ($\mathcal{PT}$-exact) or imaginary axes ($\mathcal{PT}$-broken). Depending on the spectral property, the BZ can be classified into two distinct regions: space-like and time-like regions. The momenta with \textit{space-like} spinors ($n^2_x(\textbf{k}) + n^2_z(\textbf{k}) - m^2_y(\textbf{k}) > 0$) and \textit{time-like} spinors ($n^2_x(\textbf{k}) + n^2_z(\textbf{k}) - m^2_y(\textbf{k}) < 0$) correspond to the $\mathcal{PT}$-exact and $\mathcal{PT}$-broken phases respectively \cite{realHamiltonian}. The boundary between the two region is described by \textit{light-like} spinor ($n^2_x(\textbf{k}) + n^2_z(\textbf{k}) - m^2_y(\textbf{k}) = 0$) that exhibit the EP. On the Bloch sphere, the light-like spinors are composed of two separable rings of the future and past light cones [See Fig.~\ref{fig1}(a)]. 

\cyan{Three-dimensional Hopf metal--} To exemplify the three-dimensional non-Hermitian Hopf bundle, we consider the three-component vector $\big( n_x(\mathbf{k}),m_y(\mathbf{k}),n_z(\mathbf{k}) \big)$. with the dispersion given as, $N_1(\bold k) = \sin(k_x), \ N_2(\bold k) = \sin(k_y), \ N_3(\bold k) = \sin(k_z)$ and $N_4(\bold k) = m - \cos(k_x) - \cos(k_y) -\cos(k_z)$. The Hopf invariant $w$ has a nontrivial value of +1 (-2) when $1<|m|<3$ ($-1<m<1$). Otherwise, it has trivial value \cite{HopfDuan}. Fig.~\ref{fig1}(b)-(d) shows the exemplified exceptional surfaces (green and yellow surfaces) with the non-trivial Hopf invariant, topological phase transition, and the trivial phase respectively. The interior and the exterior of the exceptional surfaces correspond to the time-like and the space-like regions respectively. The non-trivial Hopf invariant $(\omega (n_x,m_y,n_z)\neq 0)$ ensures that there exists a non-empty preimage in the BZ for any point in $S^2$. Due to the non-zero Hopf invariant, the preimage of the furture and past light-like spinor manifests as the two intertwined surfaces of the exceptional point (exceptional surface) [Fig.~\ref{fig1}(b)]. As a result, the non-trivial Hopf invariant manifests as the topologically protected two-dimensional surface of the exceptional point, which we refer to as the \textit{Hopf metal} phase. In the case of the trivial Hopf invariant [Fig. \ref{fig1}(d)], we still find the exceptional surfaces. However, these surfaces are accidental, since they can be self-annihilated by adiabatic deformation.

\begin{figure}[t!]
    \centering
    \includegraphics[width=1\linewidth]{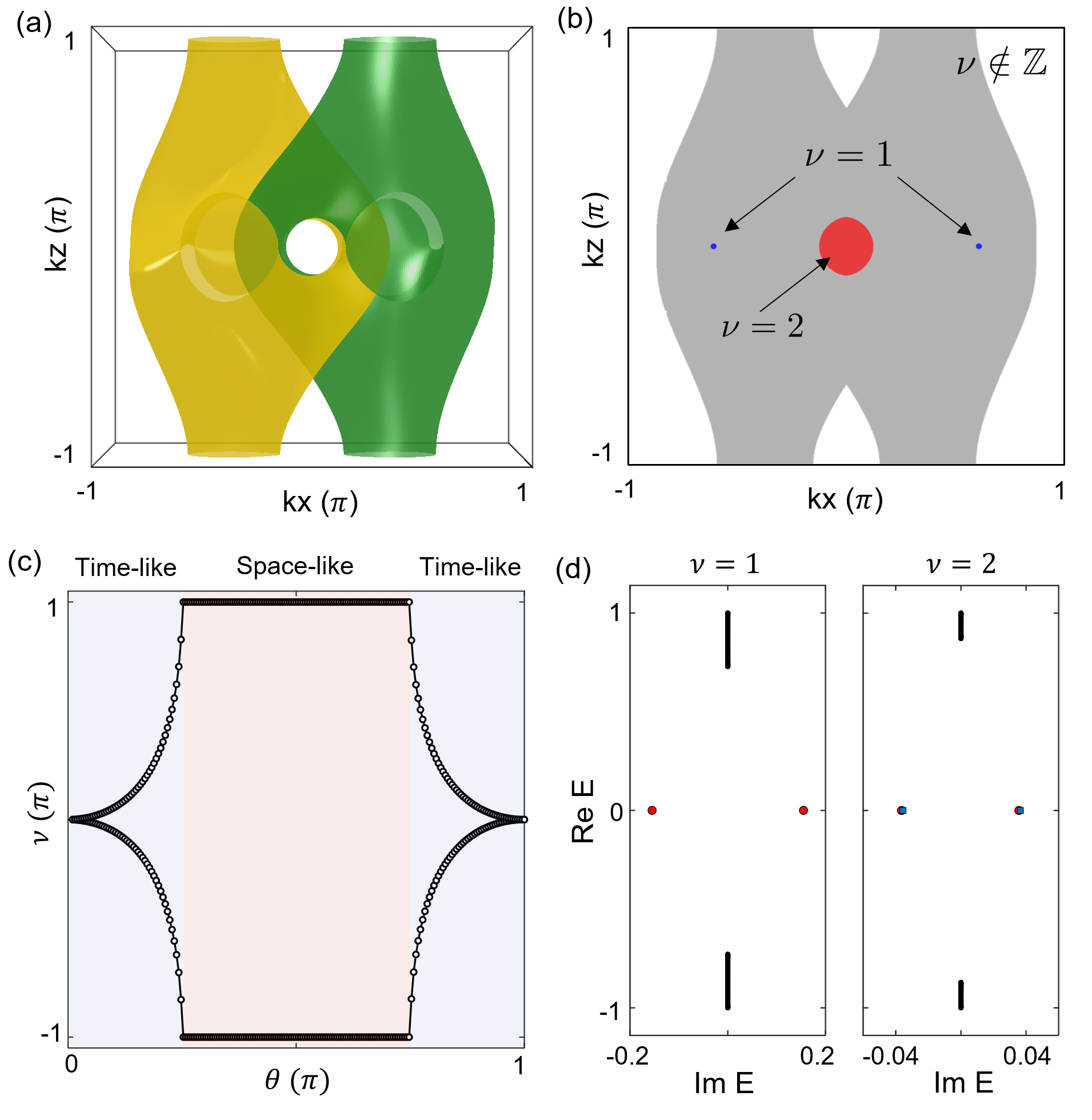} 
    \caption{(a) Projected three-dimensional Hopf metal on $k_x$-$k_z$ plane. (b) Illustration of the Zak phase integrated along $k_y$-direction. Red and blue region has non-trivial Zak phase $2\pi$ ($\nu=2$) and $\pi$ ($\nu=1$). Grey region has an ill-defined Zak phase since the line of the integration pass through the exceptional surface. (c) Calculated Zak phase as a function of polar angle on Bloch sphere. A loop of the space-like region with a winding number has quantized Zak phase of $\pi$ regardless of the presence of the non-Hermitian term. Zak phase of the loop on a time-like region loses the quantization of the Zak phase. (d) The eigenenergy spectra in the open boundary condition. Red and blue dots indicate the topological boundary modes, while the black dots represent the bulk spectra. }
    \label{fig2}
\end{figure}

\cyan{Bulk-boundary correspondence --} The exceptional surface divides the BZ into space-like and time-like regions. We can consider one-dimensional lines of the gapped regions to define the Zak phase \cite{PhysRevB.82.115120,PhysRevLett.123.066405} [For example, see red and blue regions in Fig.\ref{fig2}(b)]. In general. unlike the Hermitian systems, the presence of $\mathcal{PT}$-symmetry alone does not guarantee the quantization of the Zak phase. Nevertheless, we explicitly show that the Wilson loop of the space-like region still retains the quantization of the Zak phase [See Fig.\ref{fig2}(c)], while the Zak phase in the time-like region loses its quantization. (See supplementary material for detailed proof). On the Bloch sphere, the quantization of the Zak phase corresponds to the winding number of the loop of the spinor around the north pole. Any loop of space-like spinor has a well-defined winding number since the winding number cannot change without passing through light-like regions (EP). On the other hand, any loop of time-like spinor can adiabatically deform and it vanishes. 

As a result, we can define a quantized Zak phase $(\nu\in \mathbb{Z})$ on the exterior of the exceptional surface [Red and blue regions in Fig.\ref{fig2}(b)]. Along the one-dimensional line outside the exceptional surface, $\mathbb{Z}$-valued Zak phase can be defined as, 
\bea
\nu= \frac{1}{ i \pi } \textrm{log} \, \textrm{det} [P \textrm{exp}(\oint \mathbf{\mathcal{A}}(\mathbf{k})\cdot d\mathbf{k})],
\eea
where $\mathbf{\mathcal{A}}(\mathbf{k})=i\langle n_{R}(\mathbf{k})| \partial_\mathbf{k}| n_{R}(\mathbf{k}) \rangle$ is the non-Hermitian Berry connection. $| n_{R}(\mathbf{k}) \rangle$ is the  right eigenstate at the momentum $\mathbf{k}$. $P$ indicates the path ordering in the contour integration. Here, the Zak phase is rather $\mathbb{Z}$-valued than $\mathbb{Z}_2$ since we consider the two band model. The physical manifestation of the non-trivial Zak phase is $2\nu$ number of the topological boundary modes within the line gap in the complex energy plane [Fig.\ref{fig2}(d)]. In the three-dimensional systems with the open boundary condition along $y$-direction, the boundary modes manifests as the drumhead-like topological surface state. The drumhead surface extends as a function of $k_x$ and $k_z$ until the Wilson line touches the bulk exceptional surfaces [grey region in Fig.\ref{fig2}(b)].

\begin{figure*}[t!]
    \centering
\includegraphics[width=1\linewidth]{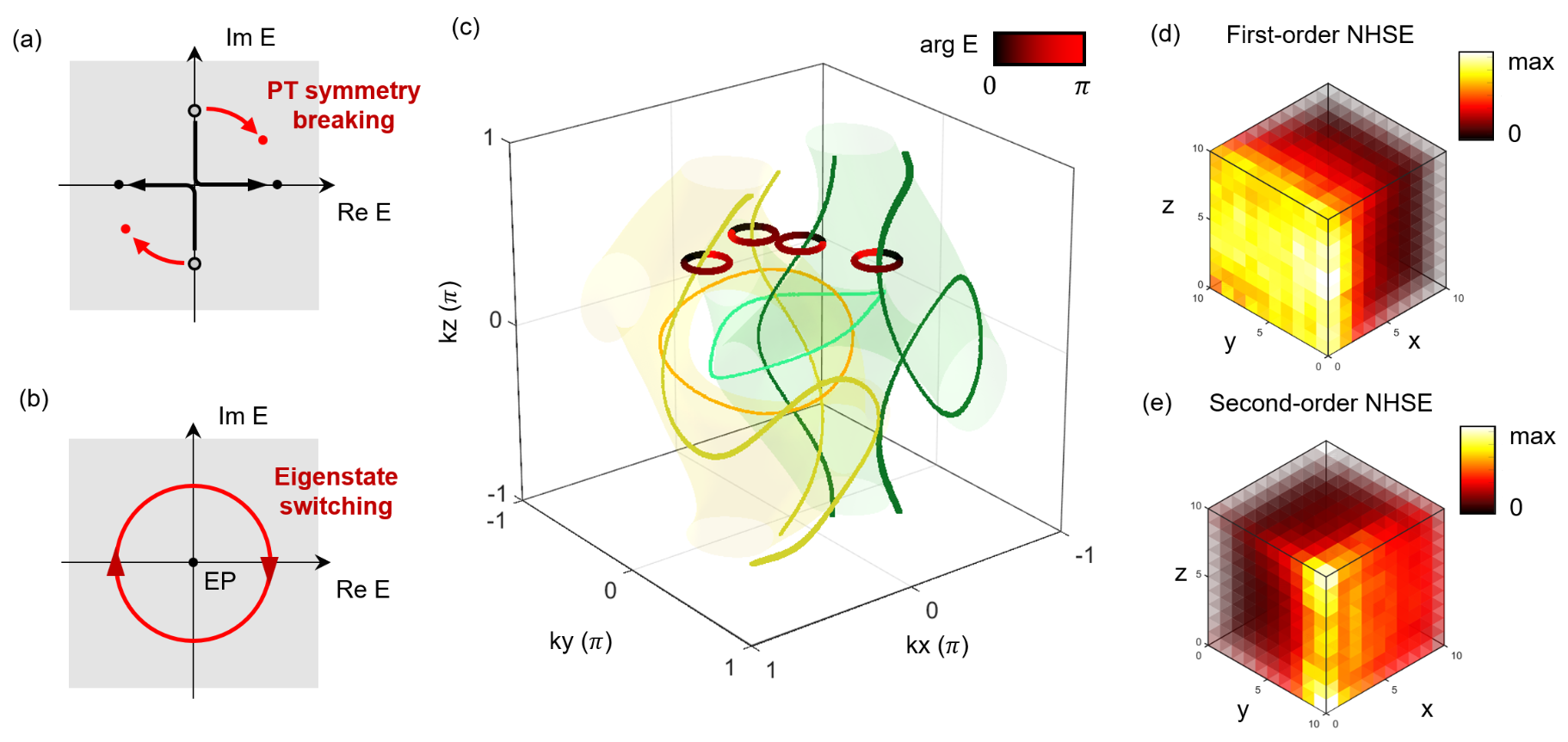}
    \caption{(a)-(b) Illustration of $\mathcal{PT}$-symmetry breaking perturbation. The presence of the perturbation results in the complex energy spectra. We can define the winding number of the complex energy as a topological invariant. (c) Deformation of the exceptional surface to exceptional nodal lines due to the $\mathcal{PT}$ symmetry-breaking perturbation. The deformed nodal lines are topologically protected by the winding of the complex energy. (d)-(e) Real-space wave functions in the open boundary conditions. By varying the symmetry-breaking perturbation, the first-order and second-order non-Hermitian skin effect is observed due to the finite spectral area in the complex energy plane.
}
\label{fig3}
\end{figure*}

\cyan{PT-symmetry broken phases--} We consider the effect of the $\mathcal{PT}$-symmetry breaking perturbation. Due to the perturbation, the energy spectra can attain an arbitrary complex phase  [See Fig.\ref{fig3}(a)]. Accordingly, the exceptional surface loses topological protection. Neverthless, the exceptional surface can deform into the topologically protected line of the EP (exceptional line) rather than the immediate line gap opening. The exceptional line has a closed line gap, and it is protected by the point gap topology  (A-class in non-Hermitian Altland-Zirnbauer classification \cite{PhysRevX.9.041015} ). We can define the vorticity ($2W\in\mathbb{Z} $) of the complex eigenvalues as,
\bea
W(E_\pm(\mathbf{k}))=\frac{1}{2\pi i}\oint d\mathbf{k} \cdot \frac{d}{d\mathbf{k}} \textrm{log} \, \textrm{det} (E_\pm(\mathbf{k})).
\eea
Here, the integration is performed over the loop that encircles the exceptional line. $E_\pm(\mathbf{k})$ is the complex eigenenergy. The integer-valued vorticity counts the number of the eigenenergy encircling ($W=2N$, $N\in \mathbb{Z}$) around the origin in the complex energy plane. The half integer-valued vorticity ($W=2N+1$) can also occur by having the eigenstate switching effect \cite{Heiss1999,heiss_sannino_1990}.

To explicitly show the formation of the exceptional line, we consider the additional anti- $\mathcal{PT}$-symmetric perturbation as, 
$
    V = 
     i\lambda_x \sigma_x + i\lambda_z \sigma_z + \lambda_y \sigma_y
    \label{eq:PTBhamiltonian}
$
where $\lambda_{x,y,z}$ are small real parameters. The corresponding perturbed energy eigenvalues are given as, $ E(\textbf{k}) = \pm[|n(\textbf{k})|^2 -|m(\textbf{k})|^2 -(\lambda_x^2 +\lambda_z^2-\lambda_y^2)  + 2i(\lambda_x n_x + \lambda_y m_y + \lambda_z n_z)]^{1\over 2} $. Then, the locations of the EP as  function of $(n_x,m_y,n_z)$ are determined by the intersections of the three surfaces, (i) sphere: $n_x^2 + m_y^2 + n_z^2 = 1$, (ii) hyperboloid: $n_x^2+n_z^2-m_y^2 -(\lambda_x^2 +\lambda_z^2-\lambda_y^2) = 0$, (iii) plain: $\lambda_x n_x + \lambda_y m_y + \lambda_z n_z = 0$. The intersection between the hyperboloid and sphere corresponds to the circles of the light-like spinors. As the intersection of the circles with the plain deforms the circles into the four distinct points, the preimage of these four points manifests as the quartet of the exceptional lines. Fig.\ref{fig3}(c) shows the examples of the exceptional lines with $W=\pm 1/2$. The exceptional lines are linked to each other as a reminiscence of the non-trivial Hopf invariant. We note that the linked structure of the exceptional lines has been previously proposed as knotted non-Hermitian metal\cite{PhysRevB.99.161115}. Hopf metal corresponds to $\mathcal{PT}$-symmetric point of the knotted non-Hermitian metal phase.

The finite spectral area due to the winding number manifests as the non-Hermitian skin effect (NHSE) in real space with the open boundary condition [Fig. \ref{fig3} (d),(e)] \cite{PhysRevLett.124.086801,PhysRevLett.125.126402,UniversalSkinChenFang}. The charge accumulation of the skin state generally occurs in one of the boundary surfaces (first-order skin effect) [Fig. \ref{fig3} (d)]. As anticipated by Zhang et al. \cite{UniversalSkinChenFang}, the control of the symmetry breaking orientation $\mathbf{\lambda}$ can induce the non-Hermitian skin effect that is dependent on the geometry of the system. By varying symmetry-breaking orientations, we also observe that the higher-order skin effect can be induced where the charge accumulations occur at the one-dimensional hinges  [Fig. \ref{fig3} (e)].

\cyan{Discussions--} In this work, we have discovered the Hopf metal phase protected by Hopf invariant in non-Hermitian $\mathcal{PT}$-symmetric systems. The topologically protected exceptional surface serves as a hallmark of the Hopf metal, and it features a direct contrast to the Hopf insulator phase in Hermitian systems. It has been known that the realization of the Hopf insulator requires long-range couplings in real space. The experimental realization of the non-Hermitian Hopf metal also demands such long-range coupling. The recent completion of the Hopf insulator in the electric circuit platform can be promising as it can efficiently design the model Hamiltonian in addition to the non-Hermitian circuit components \cite{PhysRevLett.130.057201}. The Hopf insulator has been also proposed in other physical platforms such as quantum simulator \cite{Yuan_2017} and interacting spin systems \cite{PhysRevLett.127.015301,PhysRevA.103.063322}. We also point out that both of these platforms can effectively engineer the non-reciprocal coupling\cite{PhysRevLett.127.090501,PhysRevLett.130.163001}, which promises a possible realization of the non-Hermitian Hopf metal phase.

\textit{Note} : We note that there was a proposal of non-Hermitian Hopf bundle structure in the circuit systems\cite{Kim2023}. The proposed Hopf bundle in Ref. \cite{Kim2023} shows the three-dimensional deformation of the two-dimensional insulating systems without metallic features. In Ref \cite{Kim2023}, the Hopf bundle could be defined without symmetry protection using the control of the non-Hermitian couplings on hand. In this work, we propose that $\mathcal{PT}$-symmetry systematically protects the non-trivial Hopf metal phase and the topological exceptional surface.

\section{Acknowledgement}
This work was supported by the National Research Foundation of Korea
(NRF) grant funded by the Korea government (MSIT) (Grants No. RS-2023-00252085 and No. RS-2023-00218998). S.V. and M.J.P. acknowledge financial support from the Institute for Basic Science in the Republic of Korea through the project IBS-R024-D1.

\bibliography{NHHM}

\pagebreak
\newpage

\end{document}